\documentclass[aps,twocolumn,showpacs,prl,floatfix,superscriptaddress]{revtex4}

\usepackage{epsfig}
\usepackage{graphics}
\usepackage{float}
\usepackage{lipsum}

\usepackage[T1]{fontenc}
\usepackage{mathptmx}

\begin{document}


\title{Design and Synthesis of Clathrate LaB$_{8}$ with Superconductivity}

\author{Liang Ma}
\thanks{L.M., X.Y. and G.L. equally contributed to this work.}
\affiliation{State Key Laboratory of Superhard Materials, College of Physics, Jilin University, Changchun 130012, China}
\affiliation{International Center of Computational Method \& Software, College of Physics, Jilin University, Changchun 130012, China}
\affiliation{International Center of Future Science, Jilin University, Changchun 130012, China}

\author{Xin Yang}
\thanks{L.M., X.Y. and G.L. equally contributed to this work.}
\affiliation{State Key Laboratory of Superhard Materials, College of Physics, Jilin University, Changchun 130012, China}
\affiliation{International Center of Computational Method \& Software, College of Physics, Jilin University, Changchun 130012, China}

\author{Guangtao Liu}
\thanks{L.M., X.Y. and G.L. equally contributed to this work.}
\affiliation{International Center of Computational Method \& Software, College of Physics, Jilin University, Changchun 130012, China}

\author{Hanyu Liu}
\affiliation{International Center of Computational Method \& Software, College of Physics, Jilin University, Changchun 130012, China}
\affiliation{State Key Laboratory of Superhard Materials and Key Laboratory of Physics and Technology for Advanced Batteries (Ministry of Education), College of Physics, and International Center of Future Science, Jilin University, Changchun 130012, China}

\author{Guochun Yang}
\affiliation{Centre for Advanced Optoelectronic Functional Materials Research and Key Laboratory for UV Light-Emitting Materials and Technology of Ministry of Education, Northeast Normal University, Changchun 130024, China}
\affiliation{State Key Laboratory of Metastable Materials Science \& Technology and Key Laboratory for Microstructural Material Physics of Hebei Province, School of Science, Yanshan University, Qinhuangdao 066004, China}

\author{Hui Wang}
\affiliation{Key Laboratory for Photonic and Electronic Bandgap Materials (Ministry of Education), School of Physics and Electronic Engineering, Harbin Normal University, Harbin 150025, China}

\author{Jinqun Cai}
\affiliation{State Key Laboratory of Superhard Materials, College of Physics, Jilin University, Changchun 130012, China}
\affiliation{International Center of Computational Method \& Software, College of Physics, Jilin University, Changchun 130012, China}

\author{Mi Zhou}
\email{mzhou@jlu.edu.cn}
\affiliation{International Center of Computational Method \& Software, College of Physics, Jilin University, Changchun 130012, China}

\author{Hongbo Wang}
\email{whb2477@jlu.edu.cn}
\affiliation{State Key Laboratory of Superhard Materials, College of Physics, Jilin University, Changchun 130012, China}
\affiliation{International Center of Computational Method \& Software, College of Physics, Jilin University, Changchun 130012, China}

\begin{abstract}
Boron-based clathrate materials, typically with three-dimensional networks of B atoms, have tunable properties through substitution of guest atoms, but the tuning of B cages themselves has not yet been developed. By combining crystal structural search with the laser-heated diamond anvil cell technique, we successfully synthesized a new B-based clathrate boride, LaB$_{8}$, at $\mathrm{\sim}$108 GPa and $\mathrm{\sim}$2100 K. The novel structure has a B-richest cage, with 26 B atoms encapsulating a single La atom. LaB$_{8}$ demonstrates phonon-mediated  superconductivity with an estimated  transition temperature of 14 K at ambient pressure, mainly originating from electron-phonon coupling of B cage. The replacement of La with alkaline earth metals can remarkably elevate the transition temperature. This work creates a prototype platform for subsequent investigation on tunable electronic properties through the choice of captured atoms.
\end{abstract}


\maketitle

\begin{center}
	\textbf{I. INTRODUCTION}
\end{center}

Elemental boron exhibits extraordinary structural and chemical complexity due to different arrangements of icosahedral B$_{12}$ cages with three-center bonds within icosahedra, and covalent two- and three-center bonds between icosahedra \cite{A1,A2,A3}. Through metal doping, a more B-rich B$_{24}$ cage can be obtained in a B-based clathrate metal (M) dodecaboride, MB$_{12}$, which has broadly tunable properties with different guest atoms \cite{A4,A5,A6}. To date, B-based clathrate structures have existed only in two types of MB$_{12}$ compound, namely \textit{I}4/\textit{mmm} ScB$_{12}$ and \textit{Fm}-3\textit{m} UB$_{12}$, which include sodalite-like B$_{24}$ cages with metal atoms at the center of each B cage \cite{A4}. With different guest atoms, MB$_{12}$ exhibits properties such as super-hardness (above 40 GPa for ZrB$_{12}$ \cite{A5} and Zr$_{0.5}$Y$_{0.5}$B$_{12}$ \cite{A6}), superconductivity (at 4.7 K for YB$_{12}$ and 5.8 K for ZrB$_{12}$ \cite{A7}), and oxidation resistance (at $\mathrm{\sim}$695$^{\circ}$C for Y$_{0.5}$Sc$_{0.5}$B$_{12}$ \cite{A8}), making them of broad interest for industrial application. Studies of B-based clathrate materials seem to have focused mainly on guest-atom substitution \cite{A6, A9}. However, the host cage may also play a vital role in determining material properties with, for example, YH$_{6}$ with H$_{24}$ cages and YH$_{9}$ with H$_{28}$ cages (the recently high-pressure synthesized clathrate superhydrides) having superconductivity at 227 K \cite{A10, A11} and 243 K \cite{A11,A12}, respectively, which are closely related to the high H-derived electron density of states at the Fermi level. The development of clathrate boride containing new B cages will create exciting opportunities for material innovation.

In classical clathrate structures, a suitable metal atomic radius is the primary requirement for accommodation in sodalite B$_{24}$ cages, with Y and Zr respectively being the largest and smallest metals possible \cite{A4}. Any slight size deviation renders the B-based clathrate structure unstable at ambient pressure \cite{A13}. Under high pressure, however, the chemical bonding and atomic radius are tuned, leading to synthesis of such as HfB$_{12}$, ThB$_{12}$, and GdB$_{12}$ that do not exist under ambient conditions \cite{A14,A15}. Among the known MB$_{12}$ structures, Th has the largest atomic radius \cite{A16}. Considering the sensitivity of clathrate stability to the radius of guest metal atom, formation of such a clathrate structure is inhibited when the atomic radius of the candidate metal atom differs significantly from that of existing clathrate borides. Furthermore, the formation of such a clathrate structure is associated with electron transfer from metal to B atom. Based on these factors, La may be a suitable metal candidate, with a slightly larger atomic radius and lower electronegativity than those of Th \cite{A16,A17}, which means that it might react with B at high pressures to form a novel clathrate structure. 

The formation of La-B compounds under ambient pressures has been extensively studied, but only LaB$_{4}$ and LaB$_{6}$ phases have been characterized reliably \cite{A18,A19}. Their structures feature three-dimensional B backbones of B$_{6}$ octahedra with La in interstitial sites. The LaB phase, comprising graphite-like B$^{-}$ layers, becomes stable above 30 GPa \cite{A20}. The behavior of La-B compounds at higher pressures remains relatively unexplored. Here we experimentally and theoretically investigated the chemistry of the binary La-B system at high pressures, finding a series La-B compounds thermodynamically stable at high pressure. A novel clathrate LaB$_{8}$ compound with B$_{26}$ cages was discovered under high-pressure conditions. Electronic property calculations indicate its metallic character with a superconducting transition temperature \textit{T$_{c}$} of 14 K at ambient pressure. LaB$_{8}$ exhibits tunable properties, with marked enhancement of \textit{T$_{c}$} through substitution of alkaline earth metals for La.

\begin{center}
	\textbf{II. METHODS}
\end{center}

Structure prediction of the lanthanum borides were performed by using a swarm intelligence structure search method, as implemented in the CALYPSO code \cite{A21,A22}, which has been validated with various known compounds \cite{A23,A24,A25}. We performed a structure prediction employing simulation cells with 1, 2, and 4 formula units (f.u.) at selected pressures of 1 atm and 50, 100 and 150 GPa. The \textit{ab initio} structural relaxations and electronic property calculations were performed in the framework of density functional theory \cite{A26,A27} within the generalized gradient approximation (GGA) \cite{A28} and frozen-core all-electron projector-augmented wave (PAW) method \cite{A29,A30}, as implemented in the VASP code \cite{A31,A32,A33}, where 5s$^{2}$5p$^{6}$6s$^{2}$5d$^{1}$ and 2s$^{2}$2p$^{1}$ electrons were taken as the valence electrons for La and B atoms, respectively. A plane-wave kinetic energy cutoff of 650 eV and appropriate Monkhorst-Pack k-meshes with grid spacing of 2$\pi$ $\times$ 0.03 {\AA}$^{-1}$ were chosen to ensure that enthalpy calculations were converged to less than 1 meV/atom. The phonon calculations were carried out using both finite displacement approach through the PHONOPY code \cite{A34}. The electron--phonon coupling calculations were carried out using Quantum-ESPRESSO (QE) package \cite{A35}. Ultrasoft pseudopotentials were used with a kinetic energy cutoff of 50 Ry. Bader's quantum theory of atoms in molecules (QTAIM) analysis was employed for the charge transfer calculation \cite{A36}.

Lanthanum metal (Alfa Aesar 99.5\%), LaB$_{6}$ (Alfa Aesar 99.5\%) and boron (Alfa Aesar, 99.99\%) were purchased commercially and used without further purification. In the cell-1, La foil with the thickness of 2  $\mu$m was sandwiched between the B plate ($\mathrm{\sim}$10 \textit{$\mu$}m) and MgO plate ($\mathrm{\sim}$1  $\mu$m). While in cell-2 and cell-3, binary (LaB$_{6}$ + B) mixtures were milled using Si$_{3}$N$_{4}$ media at 600 rpm for 1-min cycles over $\mathrm{\sim}$ 24 hours. The milled powders were pressed to thin plates (10  $\mu$m) then loaded into DAC sample chambers. The diamonds used in DACs had culets with diameter of 60-150  $\mu$m and were beveled at 8.5$^{o}$ to diameter of about 250  $\mu$m. Composite gasket consisting of a rhenium outer annulus and a cubic boron nitride (\textit{c}-BN) epoxy mixture insert was employed, and MgO plates served as thermal insulation from diamond anvils. The sample preparations were done in a glovebox with an inert Ar atmosphere and the contents of O$_{2}$ and H$_{2}$O less than 0.01 ppm. After being compressed to the certain pressure at room temperature, samples were heated to about 2100 K with a one-sided pulsed radiation from a YAG laser. The pressure values were determined from Raman spectra of diamond  \cite{A37} and the equation of state (EOS) of contacted MgO plate \cite{A38}.

\begin{figure}[!t]
	\begin{center}
		\epsfxsize=8.5cm
		\epsffile{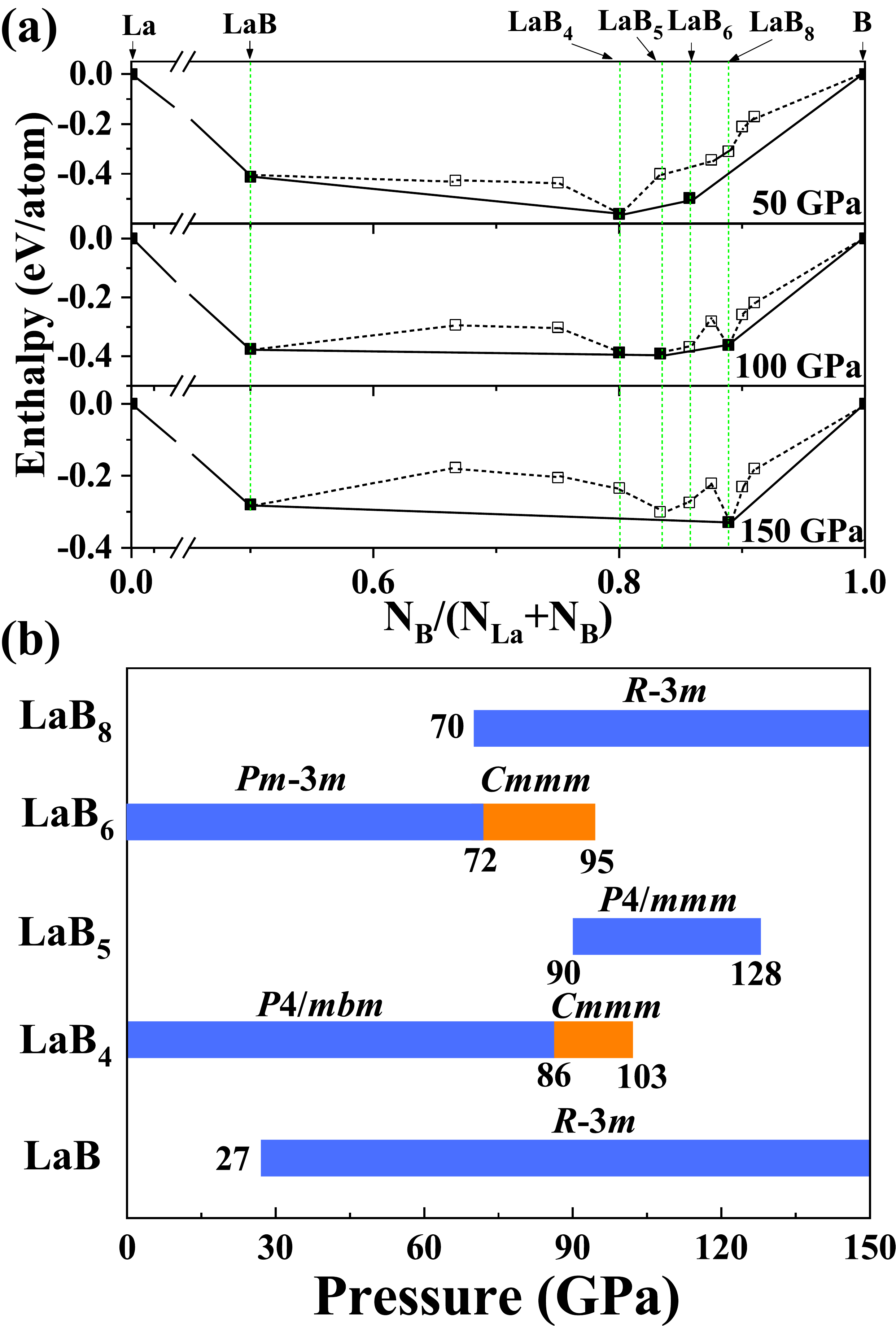}
	\end{center}
	\caption{(a) Thermodynamic stability of various La-B compounds with respect to elements La and B at selected pressures. The energetically stable phases are shown using solid symbols connected by solid lines on the convex hull. (b) Pressure-composition phase stability diagram of the La-B system.}
	\label{fig:phase}
\end{figure}

\begin{center}
	\textbf{III. RESULTS AND DISCUSSION}
\end{center}

In situ high-pressure X-ray diffraction measurements were performed at beamline BL15U1 (0.6199 {\AA}) of the Shanghai Synchrotron Radiation Facility and HP-Station 4W2 of Beijing Synchrotron Radiation Facility. Using the geometric parameters calibrated by a CeO$_{2}$ standard, the Dioptas software package was used to integrate the powder diffraction patterns then convert to 1-dimensional profiles \cite{A39}. Rietveld refinements of XRD patterns were conducted using GSAS with EXPGUI packages \cite{A40}. Additionally, the dependences of the volume on pressure were fitted by the 3rd order Birch-Murnaghan equation \cite{A41} to determine the main parameters of the EOS.

\begin{figure*}[!t]
	\begin{center}
		\epsfxsize=17cm
		\epsffile{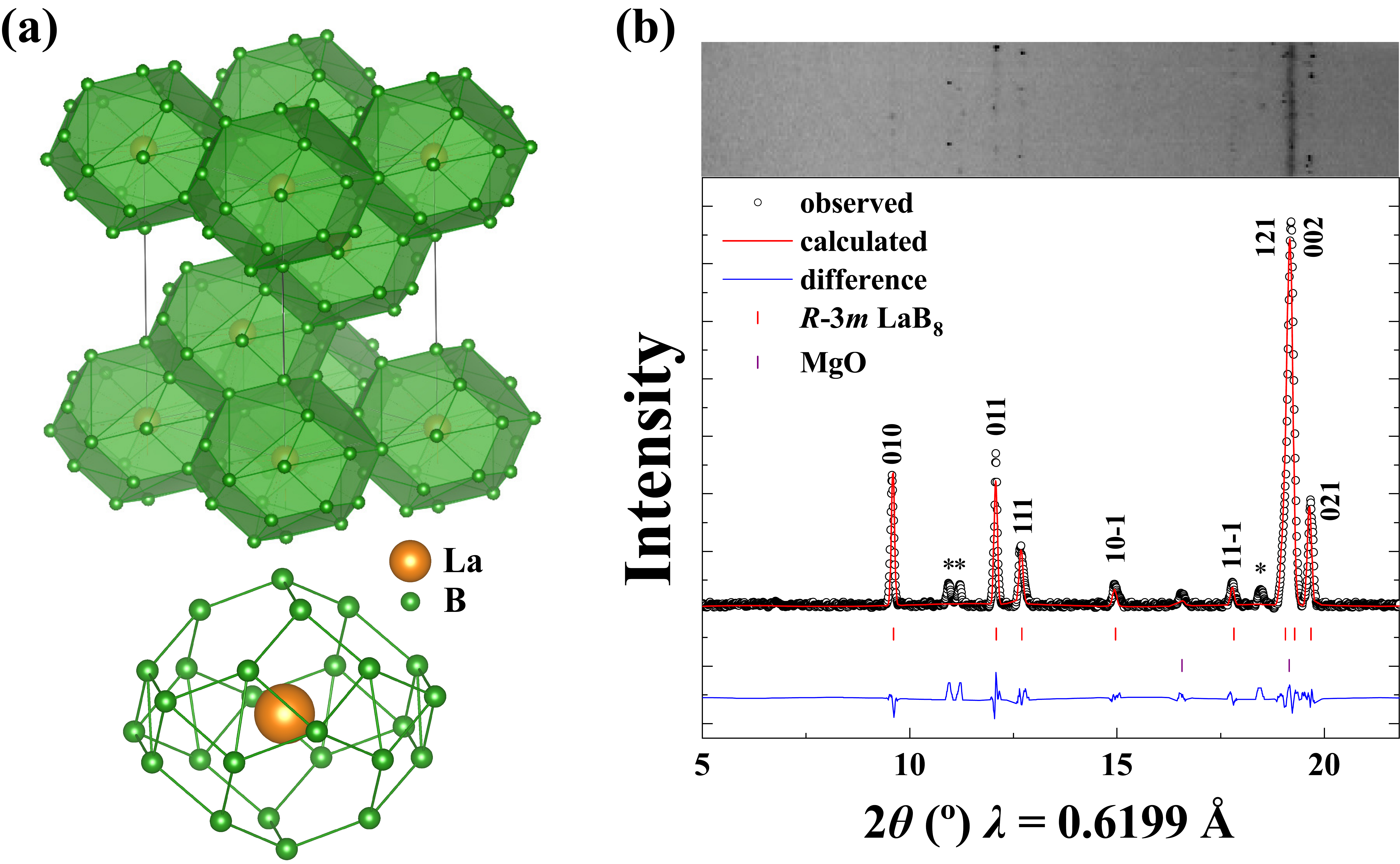}
	\end{center}
	\caption{(a) The crystal structures of clathrate LaB8. Each cage consists of twelve twisted rhombi and six twisted hexagons, with single La atom at the center. (b) Experimental XRD data (black points) of cell-3 collected at 131 GPa with Rietveld refinement (red line) of the \textit{R}-3\textit{m} LaB$_{8}$ and \textit{Fm}-3\textit{m} MgO phase. The 2D cake image is presented above the integrated pattern. Unidentified reflections are indicated by asterisks.}
	\label{fig:phase}
\end{figure*}

In seeking a new type of B-based clathrate we undertook an extensive search of known structures based on stoichiometric LaB$_{x}$ (x = 1--10) at pressures of 50, 100, and 150 GPa (Fig. 1). The energetic stabilities of LaB$_{x}$ compounds were evaluated from their formation enthalpies with respect to the dissociation products of La \cite{A42} and B \cite{A1} solids. Previously known LaB, LaB$_{4}$, and LaB$_{6}$ phases were readily identified \cite{A18,A19,A20}, indicating the reliability of our calculations (Fig. 1b). The high-pressure phases of LaB$_{4}$ and LaB$_{6}$, both with a \textit{Cmmm} space group (Fig. S1), and two unexpected compositions of LaB$_{5}$ and LaB$_{8}$ become stable at elevated pressures. Detailed structural information, electronic properties, and phonon dispersion curves are shown in Figs. S2--S4 and Table S1. 

Interestingly, LaB$_{8}$ exhibits peculiar three-dimensional B clathrate structures comprising contiguous face-sharing B$_{26}$ cages (Fig. 2a) in which single La atoms are nested. These cages are constructed from 12 twisted rhombi and 6 twisted hexagons, with B--B bond lengths of 1.65--1.72 {\AA} at 100 GPa. Enthalpy calculations reveal its thermodynamic stability above 70 GPa. LaB$_{8}$ appears to be an unprecedented stoichiometry in metal borides that contain more B-rich cages than conventional B$_{24}$ cages in clathrate MB$_{12}$. In phonon calculations, clathrate LaB$_{8}$ exhibits no imaginary phonon frequencies at 0--150 GPa (Fig. S4), indicating its dynamical stability. The electron localization function (ELF) and Bader charge analysis were undertaken to explore the bonding nature of the clathrate structure LaB$_{8}$ (Fig. S5 and Table S2). Results indicate strong covalent bonding between B atoms and weak ionic bonding between La and the B$_{26}$ cage. Because of the strong B--B covalent bonding, we investigated the specific Vickers hardness (\textit{H$_{v}$}) of \textit{R}-3\textit{m} LaB$_{8}$ using an empirical model \cite{A43}. The calculated hardness of \textit{R}-3\textit{m} LaB$_{8}$ is 25.3 GPa, indicating the possibility of its being a hard material.

Guided by theoretical prediction, we carried out laser-heated diamond anvil cell (LHDAC) experiments to synthesize the target clathrate LaB$_{8}$. In cell-1, a piece of La foil and excess B powder were compressed to 115 GPa and heated to $\mathrm{\sim}$2100 K. X-ray diffraction (XRD) patterns (Fig. S6a) indicate a complex mixture of products with predominant LaB$_{8}$ marked by a red bar below the pattern. There were also some unidentified patterns.

\begin{figure*}[!t]
	\begin{center}
		\epsfxsize=17cm
		\epsffile{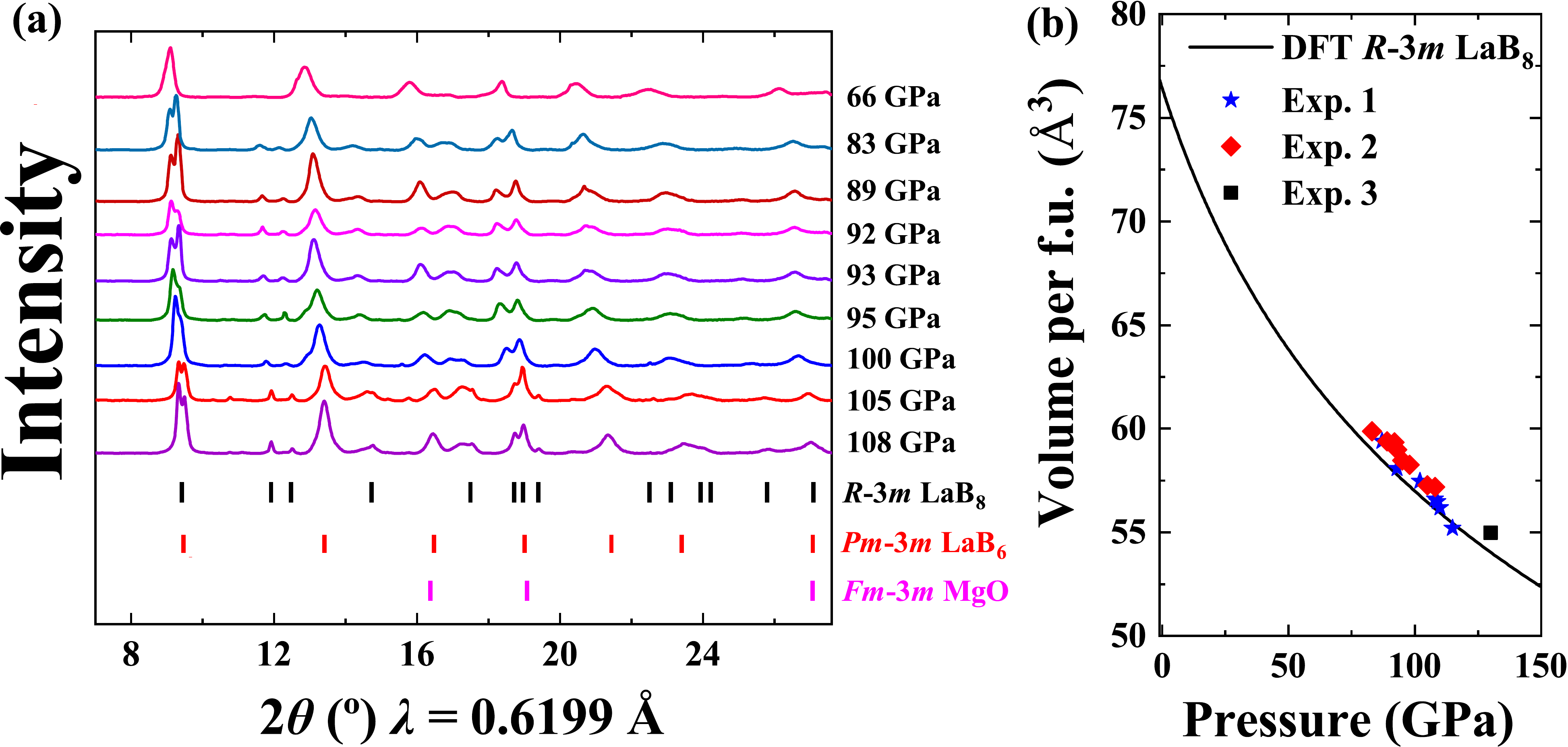}
	\end{center}
	\caption{(a) Experimental XRD patterns of cell-2 during decompression in the pressure range of 66-108 GPa. (b) Experimental EOS data for the sample are shown as points, in comparison with that of predicted \textit{R}-3\textit{m} LaB$_{8}$ (GGA PBE). The pV data were fitted using the third order Birch-Murnaghan EOS.}
	\label{fig:phase}
\end{figure*}

To obtain high-quality XRD patterns of clathrate LaB$_{8}$, homogeneous fine-grained mixtures of LaB$_{6}$ and B were used as precursors in cell-2 and cell-3 experiments, targeting the stoichiometric reaction, LaB$_{6}$ + 2B $\mathrm{\to}$ LaB$_{8}$. Samples were compressed to 108 GPa and 131 GPa, respectively, then laser heated to $\mathrm{\sim}$2100 K. XRD patterns (Figs 2b and S7) are consistent with the patterns calculated for clathrate LaB$_{8}$. The peaks marked with asterisks in Fig. 2b could not be reproduced in other experiments, indicating that they may be due to other LaB stochiometries or impurities. Upon decompression, LaB$_{8}$ was stable to at least 83 GPa, decomposing to LaB$_{6}$ and B at 66 GPa (Fig. 3a). Due to the low atomic scattering power of B, we directly determined the positions of La atoms from XRD data. Therefore, we further developed the equation of state (EOS) by decreasing the pressure in the cell-1 and cell-2 experiments, consistent with the theoretical EOS of LaB$_{8}$ (Fig. 3b).

The discovery of superconductivity in MgB$_{2}$ at 39 K \cite{A44} provided a new perspective in the search for metal boride materials with high \textit{T$_{c}$}, with strong electron--phonon coupling of B layers in the structure being largely responsible for the high \textit{T$_{c}$} \cite{A45}. However, the rather low superconductivity of 0.39 K, 4.7 K, 5.8 K and 0.4 K were observed in conventional clathrate borides MB$_{12}$, for ScB$_{12}$, YB$_{12}$, ZrB$_{12}$ and LuB$_{12}$, respectively \cite{A46,A47}. Therefore, we are quite curious about the superconductivity in the novel clathrate boride LaB${}_{8}$. We first performed the electronic property calculations for LaB$_{8}$ at selected pressures and compared the results with those for LaB$_{12}$, the structure of which was constructed from known prototype structures of \textit{I}4/\textit{mmm} ScB$_{12}$ and \textit{Fm}-3\textit{m} UB$_{12}$ (Figs 4a and S8). Results indicate the metallic nature of all three structures, with the B-derived density of state (DOS) value of LaB$_{8}$ around the Fermi level (FL) being higher than that of LaB$_{12}$, which means the superconductivity may be significantly improved. The superconducting \textit{T$_{c}$} value was estimated using the Allen-Dynes modified McMillan equation.(Figs 4b and S9). The  electron--phonon coupling constant \textit{$\lambda$} of LaB$_{8}$ is calculated to be 0.61, which is larger than that of \textit{Fm}-3\textit{m} LaB$_{12}$ (0.25) and \textit{I}4/\textit{mmm} LaB$_{12}$ (0.24), leading the estimated \textit{T$_{c}$} as high as 14 K at ambient pressure with $\mu$* of 0.1, much higher than 0.01 K and 0.01 K in \textit{Fm}-3\textit{m} LaB$_{12}$ and \textit{I}4/\textit{mmm} LaB$_{12}$. 

The phonon dispersion, projected phonon DOS, and Eliashberg spectral function (\textit{$\alpha$}$^{2}$F(\textit{$\omega$}) and its integral \textit{$\lambda$}(\textit{$\omega$})) calculations of LaB$_{8}$ were all undertaken to explore the superconducting properties (Fig. 4b). The calculated phonon DOS can be separated into two regions: low-frequency vibrations (0--6 THz) related to La atoms, and the high-frequency vibrations related to B atoms. The vibrations of B atoms in a large range of 6--26 THz contribute as much as 89\% to the total \textit{$\lambda$}. Considering the significant DOS at the FL of B atoms, we therefore conclude that the coupling between the electrons and phonon vibrations of B atoms is responsible for the high superconductivity of LaB$_{8}$. While the average phonon frequency gradually increases with decreasing \textit{$\lambda$} at higher pressures, leading to a predicted \textit{T$_{c}$} of 2.1 K at 100 GPa.

The electronic DOS of LaB$_{8}$ exhibits a peak below the FL (Fig. 4a), providing a feasible means of increasing the electronic DOS around the FL through the tunable electronic structure of the B$_{26}$ cage in LaB$_{8}$. We substituted La of \textit{R}-3\textit{m} LaB$_{8}$ with other guest ions to construct a series of \textit{R}-3\textit{m} structures of MB$_{8}$ compounds (M = Mg, Ca, Sr, Y). Phonon dispersion relations of these structures indicate their dynamic stability at ambient pressure (Fig. S4). Electron--phonon calculations of substituted MB$_{8}$ were performed at ambient pressure, with specific \textit{$\lambda$, $\alpha$}$^{2}$F(\textit{$\omega$}), and \textit{T$_{c}$} values as listed in Figs S9 and S10. With alkaline earth elements inserted into the B$_{26}$ cage there was less electron transfer to B atoms than La atom, leading to a notable increase in electron density around the FL (Fig. S10). As anticipated, the MB$_{8}$ (M = Mg, Ca, Sr) structures demonstrate high-temperature superconductivity at ambient pressure (Fig. S10). The current results could provide a theoretical guidance for future experiments.

\begin{figure}[!t]
	\begin{center}
		\epsfxsize=8.5cm
		\epsffile{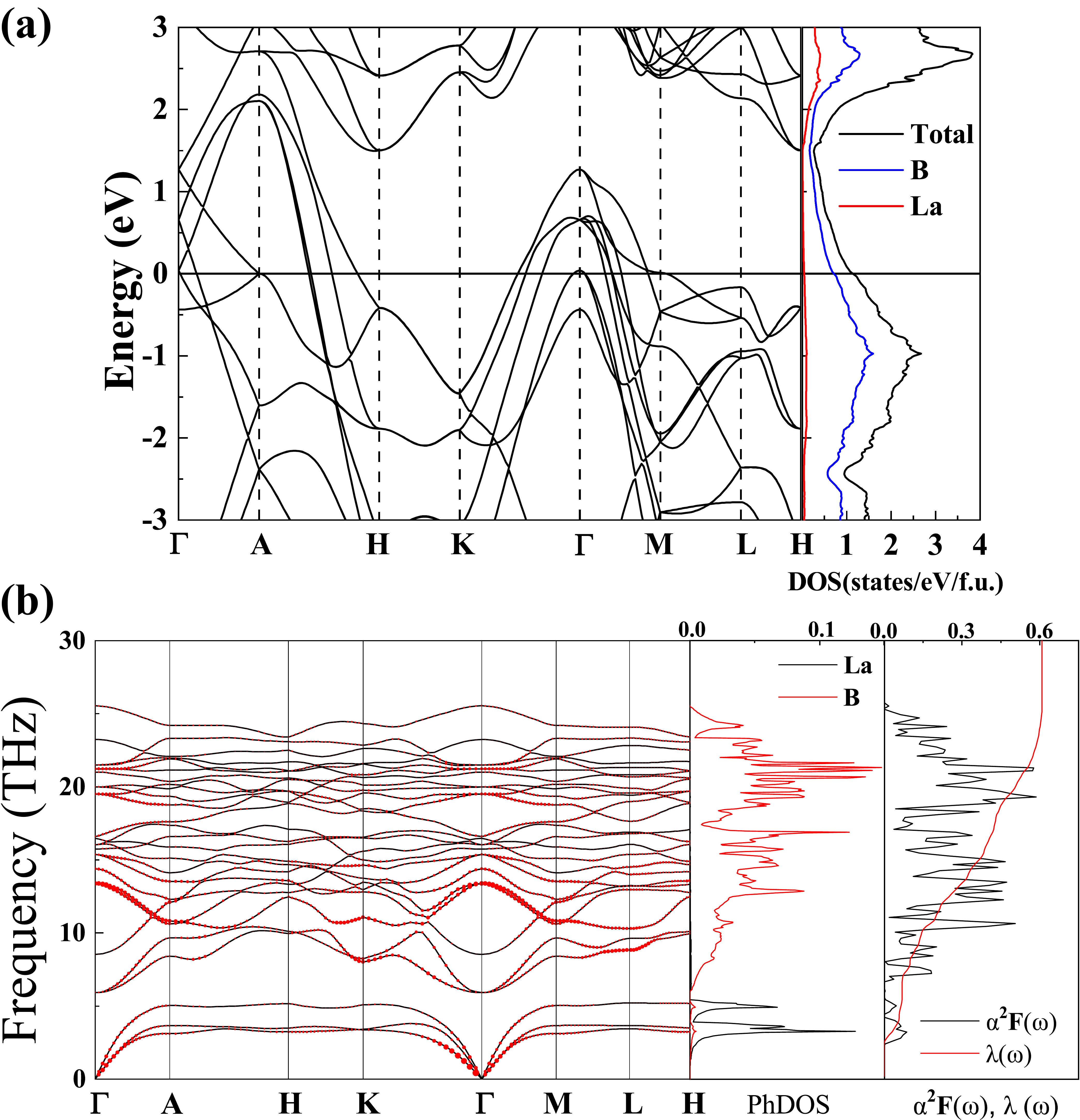}
	\end{center}
	\caption{(a) The electronic band structure and projected density of states (DOS) for \textit{R}-3\textit{m} LaB$_{8}$ at 0 GPa. (b) Phonon dispersion relations, projected phonon density of states (PHDOS) and Eliashberg spectral function for of LaB$_{8}$ at 0 GPa. The size of the red dots represents the magnitude of the EPC.}
	\label{fig:phase}
\end{figure}

\begin{center}
	\textbf{IV. CONCLUSION }
\end{center}

The high-pressure behavior of binary La-B compounds was investigated by combining a crystal structural search and the LHDAC technique. A novel clathrate boride, LaB$_{8}$, with a three-dimensional B network of B$_{26}$ cages with a La atom at the center was successfully synthesized at high pressure and temperature. Electronic property calculations indicate its metallic nature with a predicted\textit{ T$_{c}$ }of 14 K at ambient pressure. Subsequent calculations indicated its tunable properties, with \textit{T$_{c}$} values showing a great improvement with substitution of alkaline earth metals for La. This study provides a new platform for the design of materials with tunable electronic properties, and may stimulate high-pressure experimental work on the synthesis of high-\textit{T$_{c}$} superconductors based on clathrate structures.

${}$
${}$
${}$

This work was supported by the Major Program of the National Natural Science Foundation of China (Grant No. 52090024), the Strategic Priority Research Program of Chinese Academy of Sciences (Grant No. XDB33000000), National Key R\&D Program of China (Grant no. 2018YFA0305900), National Natural Science Foundation of China (Grant No. 11874175, 11974135, 12074139, 12074138, 11874176 and 12034009), Jilin Province Outstanding Young Talents Project (Grant No. 20190103040JH), and Program for JLU Science and Technology Innovative Research Team (JLUSTIRT). XRD measurement were performed at Shanghai Synchrotron Radiation Facility Beamline BL15U1, Beijing Synchrotron Radiation Facility HP-Station 4W2 and BL10XU/SPring-8.


\begin{thebibliography}{10}
\expandafter\ifx\csname natexlab\endcsname\relax\def\natexlab#1{#1}\fi
\expandafter\ifx\csname bibnamefont\endcsname\relax
  \def\bibnamefont#1{#1}\fi
\expandafter\ifx\csname bibfnamefont\endcsname\relax
  \def\bibfnamefont#1{#1}\fi
\expandafter\ifx\csname citenamefont\endcsname\relax
  \def\citenamefont#1{#1}\fi
\expandafter\ifx\csname url\endcsname\relax
  \def\url#1{\texttt{#1}}\fi
\expandafter\ifx\csname urlprefix\endcsname\relax\def\urlprefix{URL }\fi
\providecommand{\bibinfo}[2]{#2}
\providecommand{\eprint}[2][]{\url{#2}}

\bibitem{A1} A. R. Oganov\textit{ et al.}, Nature \textbf{457}, 863 (2009).

\bibitem{A2} R. E. Hughes, C. H. L. Kennard, D. B. Sullenger, H. A. Weakliem, D. E. Sands, and J. L. Hoard, J. Am. Chem. Soc. \textbf{85}, 361 (1963).

\bibitem{A3} B. Albert and H. Hillebrecht, Angew. Chem., Int. Ed. \textbf{48}, 8640 (2009).

\bibitem{A4} G. Akopov, M. T. Yeung, and R. B. Kaner, Adv. Mater. \textbf{29}, 1604506 (2017).

\bibitem{A5} T. Ma\textit{ et al.}, Adv. Mater. \textbf{29}, 1604003 (2017).

\bibitem{A6} G. Akopov\textit{ et al.}, J. Am. Chem. Soc. \textbf{141}, 9047 (2019).

\bibitem{A7} R. Tro\'{c}, R. Wawryk, A. Pikul, and N. Shitsevalova, Philos. Mag. \textbf{95}, 2343 (2015).

\bibitem{A8} Y. Liang\textit{ et al.}, Chem. Mater. \textbf{31}, 1075 (2019).

\bibitem{A9} J. Lei, G. Akopov, M. T. Yeung, J. Yan, R. B. Kaner, and S. H. Tolbert, Adv. Funct. Mater. \textbf{29}, 1900293 (2019).

\bibitem{A10} I. A. Troyan\textit{ et al.}, Adv. Mater. \textbf{33}, 2006832 (2021).

\bibitem{A11} P. P. Kong\textit{ et al.}, arXiv:1909.10482 (2019).

\bibitem{A12} E. Snider, N. Dasenbrock-Gammon, R. McBride, X. Wang, N. Meyers, K. V. Lawler, E. Zurek, A. Salamat, and R. P. Dias, Phys. Rev. Lett. \textbf{126}, 117003 (2021).

\bibitem{A13} S. La Placa, I. Binder, and B. Post, J. Inorg. Nucl. Chem. \textbf{18}, 113 (1961).

\bibitem{A14} J. F. Cannon and P. B. Farnsworth, J. Less-Common Met. \textbf{92}, 359 (1983).

\bibitem{A15} J. F. Cannon, D. M. Cannon, and H. T. Hall, J. Less-Common Met. \textbf{56}, 83 (1977).

\bibitem{A16} J. C. Slater, J. Chem. Phys. \textbf{41}, 3199 (1964).

\bibitem{A17} M. Rahm, R. Cammi, N. W. Ashcroft, and R. Hoffmann, J. Am. Chem. Soc. \textbf{141}, 10253 (2019).

\bibitem{A18} K. Kato, I. Kawada, C. Oshima, and S. Kawai, Acta Crystallogr., Sect. B: Struct. Crystallogr. Cryst. Chem. \textbf{30}, 2933 (1974).

\bibitem{A19} M. E. Schlesinger, P. K. Liao, and K. E. Spear, J. Phase Equilib. \textbf{20}, 73 (1999).

\bibitem{A20} A. G. Van Der Geest and A. N. Kolmogorov, Calphad \textbf{46}, 184 (2014).

\bibitem{A21} Y. Wang, J. Lv, L. Zhu, and Y. Ma, Phys. Rev. B \textbf{82}, 094116 (2010).

\bibitem{A22} Y. Wang, J. Lv, L. Zhu, and Y. Ma, Comput. Phys. Commun. \textbf{183}, 2063 (2012).

\bibitem{A23} X. Yang, H. Li, H. Liu, H. Wang, Y. Yao, and Y. Xie, Phys. Rev. B \textbf{101}, 184113 (2020).

\bibitem{A24} F. Peng, Y. Sun, C. J. Pickard, R. J. Needs, Q. Wu, and Y. Ma, Phys. Rev. Lett. \textbf{119}, 107001 (2017).

\bibitem{A25} L. Ma, G. Liu, Y. Wang, M. Zhou, H. Liu, F. Peng, H. Wang, and Y. Ma, arXiv:2002.09900 (2020). 

\bibitem{A26} P. Hohenberg and W. Kohn, Phys. Rev. \textbf{136}, B864 (1964).

\bibitem{A27} W. Kohn and L. J. Sham, Phys. Rev. \textbf{140}, A1133 (1965).

\bibitem{A28} J. P. Perdew, K. Burke, and M. Ernzerhof, Phys. Rev. Lett. \textbf{77}, 3865 (1996).

\bibitem{A29} P. E. Bl?chl, Phys. Rev. B \textbf{50}, 17953 (1994).

\bibitem{A30} G. Kresse and D. Joubert, Phys. Rev. B \textbf{59}, 1758 (1999).

\bibitem{A31} G. Kresse and J. Hafner, Phys. Rev. B \textbf{48}, 13115 (1993).

\bibitem{A32} G. Kresse and J. Hafner, Phys. Rev. B \textbf{49}, 14251 (1994).

\bibitem{A33} G. Kresse and J. Furthm?ller, Comput. Mater. Sci. \textbf{6}, 15 (1996).

\bibitem{A34} A. Togo, F. Oba, and I. Tanaka, Phys. Rev. B \textbf{78}, 134106 (2008).

\bibitem{A35} P. Giannozzi\textit{ et al.}, J. Phys.: Condens. Matter \textbf{21}, 395502 (2009).

\bibitem{A36} W. Tang, E. Sanville, and G. Henkelman, J. Phys.: Condens. Matter \textbf{21}, 084204 (2009).

\bibitem{A37} Y. Akahama and H. Kawamura, J. Appl. Phys. \textbf{100}, 043516 (2006).

\bibitem{A38} S. M. Dorfman, V. B. Prakapenka, Y. Meng, and T. S. Duffy, J. Geophys. Res.: Solid Earth \textbf{117}, B08210 (2012).

\bibitem{A39} C. Prescher and V. B. Prakapenka, High Pressure Res. \textbf{35}, 223 (2015).

\bibitem{A40} B. H. Toby, J. Appl. Crystallogr. \textbf{34}, 210 (2001).

\bibitem{A41} F. Birch, Phys. Rev. \textbf{71}, 809 (1947).

\bibitem{A42} W. Chen, D. V. Semenok, I. A. Troyan, A. G. Ivanova, X. Huang, A. R. Oganov, and T. Cui, Phys. Rev. B \textbf{102}, 134510 (2020).

\bibitem{A43} J. S. Tse, J. Super. Mater. \textbf{32}, 177 (2010).

\bibitem{A44} J. Nagamatsu, N. Nakagawa, T. Muranaka, Y. Zenitani, and J. Akimitsu, Nature \textbf{410}, 63 (2001).

\bibitem{A45} J. M. An and W. E. Pickett, Phys. Rev. Lett. \textbf{86}, 4366 (2001).

\bibitem{A46} S. T. Matthias, T. H. Geballe, K. Andres, E. Corenzwit, G. W. Hull and J. P. Maita, Science \textbf{159}, 530 (1968).

\bibitem{A47} I. Bat'ko, M. Bat'kov'a, K. Flachbart, V. B. Filippov, Y. B. Paderno, N. Y. Shitsevalova and T. Wagner, J. Alloys Compd. \textbf{217}, L1 (1995).

\end{thebibliography}
\end{document}


\title{Supplementary Material for\\ ``Design and Synthesis of Clathrate LaB$_{8}$ with Superconductivity''}

\author{{ Liang Ma,$^{1,2,3,\star}$ Xin Yang,$^{1,2,\star}$ Guangtao Liu,$^{2,\star}$ Hanyu Liu,$^{2,4}$ Guochun Yang,$^{5,6}$ Hui Wang,$^{7}$ Jinqun Cai,$^{1,2}$  Mi Zhou,$^{2,\dagger}$ and Hongbo Wang,$^{1,2,\ddagger}$} \\
{\small \em $^1$State Key Laboratory of Superhard Materials, College of Physics, Jilin University, Changchun 130012, China\\
$^2$International Center of Computational Method \& Software, College of Physics, Jilin University, Changchun 130012, China\\
$^3$International Center of Future Science, Jilin University, Changchun 130012, China\\
$^4$4State Key Laboratory of Superhard Materials and Key Laboratory of Physics and Technology for Advanced Batteries (Ministry of Education), College of Physics, and International Center of Future Science, Jilin University, Changchun 130012, China\\
$^5$Centre for Advanced Optoelectronic Functional Materials Research and Key Laboratory for UV Light-Emitting Materials and Technology of Ministry of Education, Northeast Normal University, Changchun 130024, China\\
$^6$State Key Laboratory of Metastable Materials Science \& Technology and Key Laboratory for Microstructural Material Physics of Hebei Province, School of Science, Yanshan University, Qinhuangdao 066004, China\\
$^7$Key Laboratory for Photonic and Electronic Bandgap Materials (Ministry of Education), School of Physics and Electronic Engineering, Harbin Normal University, Harbin 150025, China}
{\small $^{\dagger}$Electronic Address: mzhou@jlu.edu.cn\\
$^{\ddagger}$Electronic Address: whb2477@jlu.edu.cn\\
$^{\star}$ L.M., Y.X. and G.L. equally contributed to this work}}

\date{\today}

\maketitle

\begin{center}
	\epsfxsize=14cm
	\epsffile{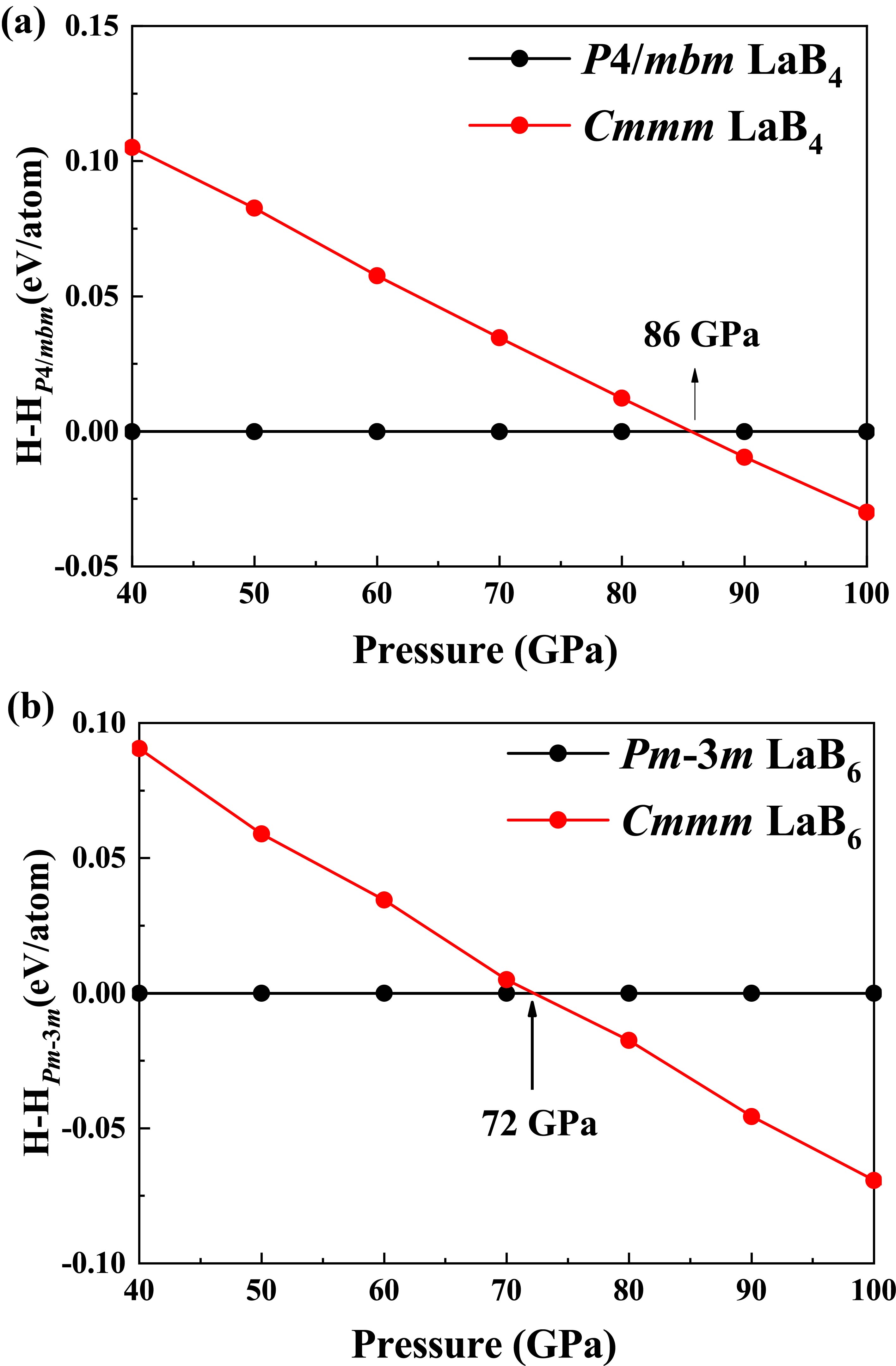}
\end{center}
\textbf{Fig. S1.} Fig. S1. (a) Relative stability of \textit{Cmmm} LaB$_{6}$ with respect to \textit{Pm}-3\textit{m} LaB$_{6}$. (b) Relative stability of \textit{Cmmm} LaB$_{4}$ with respect to \textit{P}4/\textit{mbm} LaB$_{4}$.
\label{fig:phase}

\newpage
\begin{center}
	\epsfxsize=14cm
	\epsffile{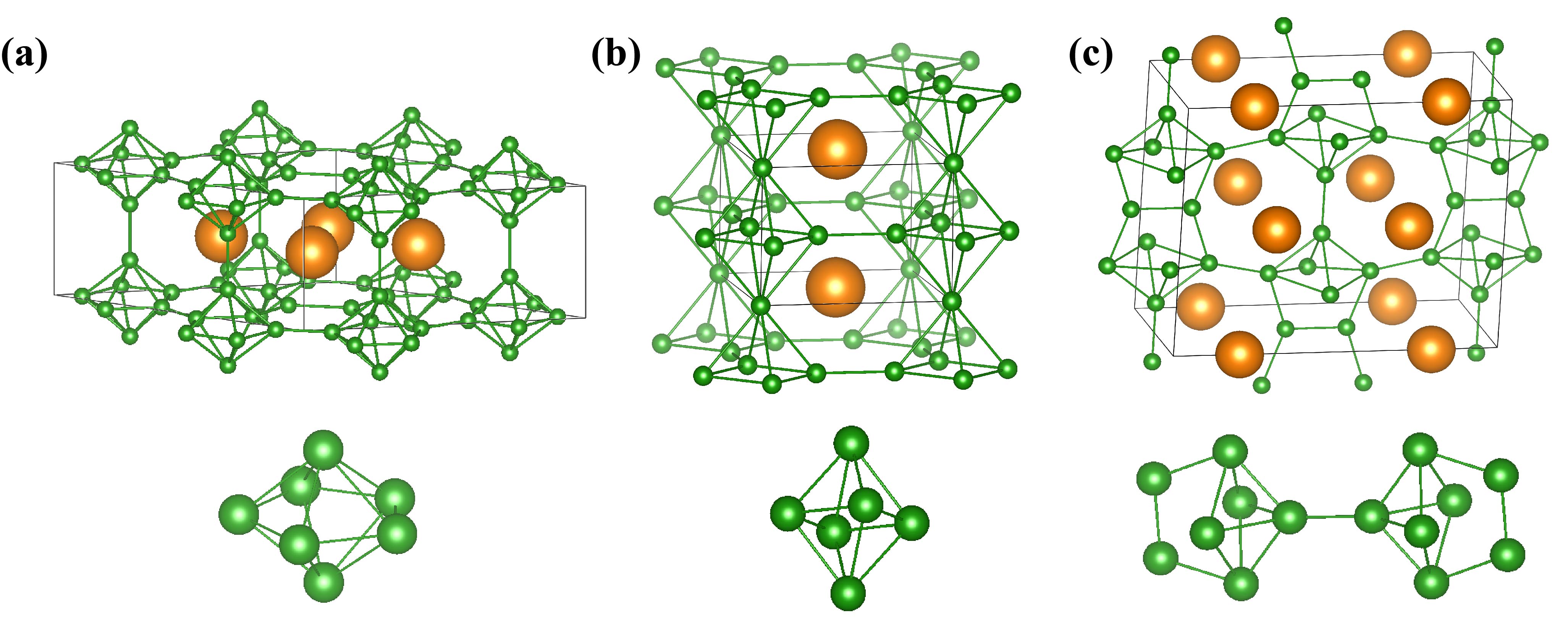}
\end{center}	
\textbf{FIG. S2.} The predicted stable crystal structures: (a)\textit{Cmmm} LaB$_{4}$ structure, (b) \textit{P}4/\textit{mmm} LaB$_{5}$ structure, (c) \textit{Cmmm} LaB$_{6}$ structure. Large brown and small green spheres represent La and B atoms, respectively. The \textit{Cmmm} LaB$_{4}$ phase has an orthorhombic structure, consisting of B$_{7}$ pentagonal bipyramids, where the La atoms sitting in the interstitial sites. While the \textit{P}4/\textit{mmm} LaB$_{5}$ phase stabilizes into a tetragonal structure, which is composed of B$_{6}$ octahedra, forming an open-channel frame with the La atoms embedded. \textit{Cmmm} LaB$_{6}$ phase contains a three-dimensional boron network with twinned B$_{7}$ pentagonal bipyramids units interconnected and the La atoms locate in the skeletal channels.
\label{fig:phase}

\begin{center}
	\epsfxsize=14cm
	\epsffile{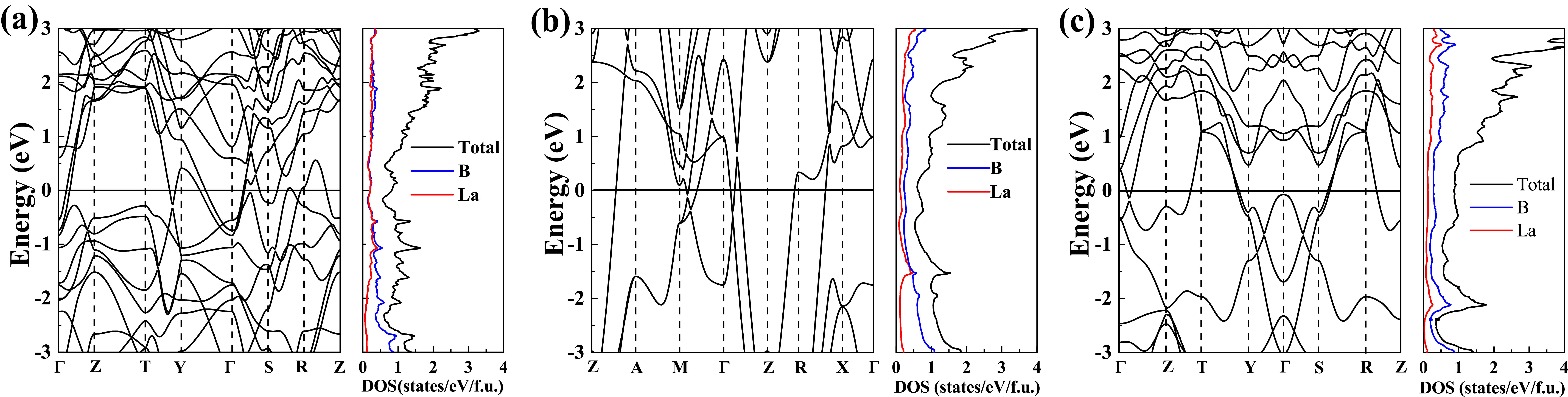}
\end{center}	
\textbf{FIG. S3.} Electronic band structures, corresponding projected density of states (PDOS) and phonon dispersion curves of the predicted La-B compounds at selected pressures. (a) \textit{Cmmm} LaB$_{4}$ at 100 GPa (b) \textit{P}4/\textit{mmm} LaB$_{5}$ at 80 GPa and (c) \textit{Cmmm} LaB$_{6}$ at 75 GPa. 
\label{fig:phase}

\newpage
\begin{center}
	\epsfxsize=12cm
	\epsffile{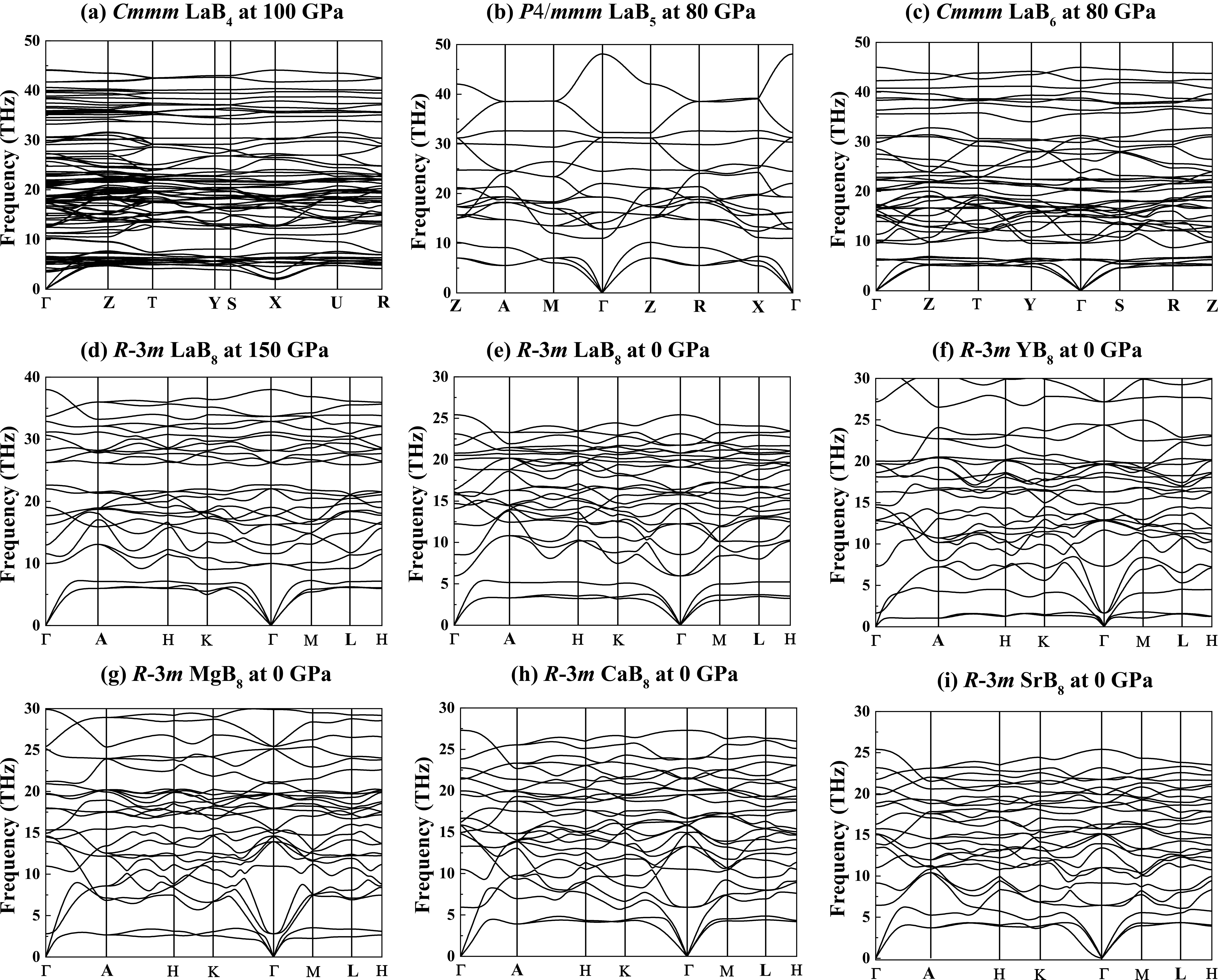}
\end{center}	
\textbf{FIG. S4.} Phonon dispersion curves of the predicted Lanthanum borides (a) \textit{Cmmm} LaB$_{4}$ at 100 GPa (b) \textit{P}4/\textit{mmm} LaB$_{5}$ at 80 GPa (c)\textit{Cmmm} LaB$_{6}$ at 80 GPa. (d) \textit{R}-3\textit{m} LaB$_{8}$ at 150 GPa (e) \textit{R}-3\textit{m} LaB$_{8}$ at 0 GPa (f) \textit{R}-3\textit{m} YB$_{8}$ at 0 GPa (g)\textit{R}-3\textit{m} MgB$_{8}$ at 0 GPa (h) \textit{R}-3\textit{m} CaB$_{8}$ at 0 GPa (i) \textit{R}-3\textit{m} SrB$_{8}$ at 0 GPa. 
\label{fig:phase}

\begin{center}
	\epsfxsize=12cm
	\epsffile{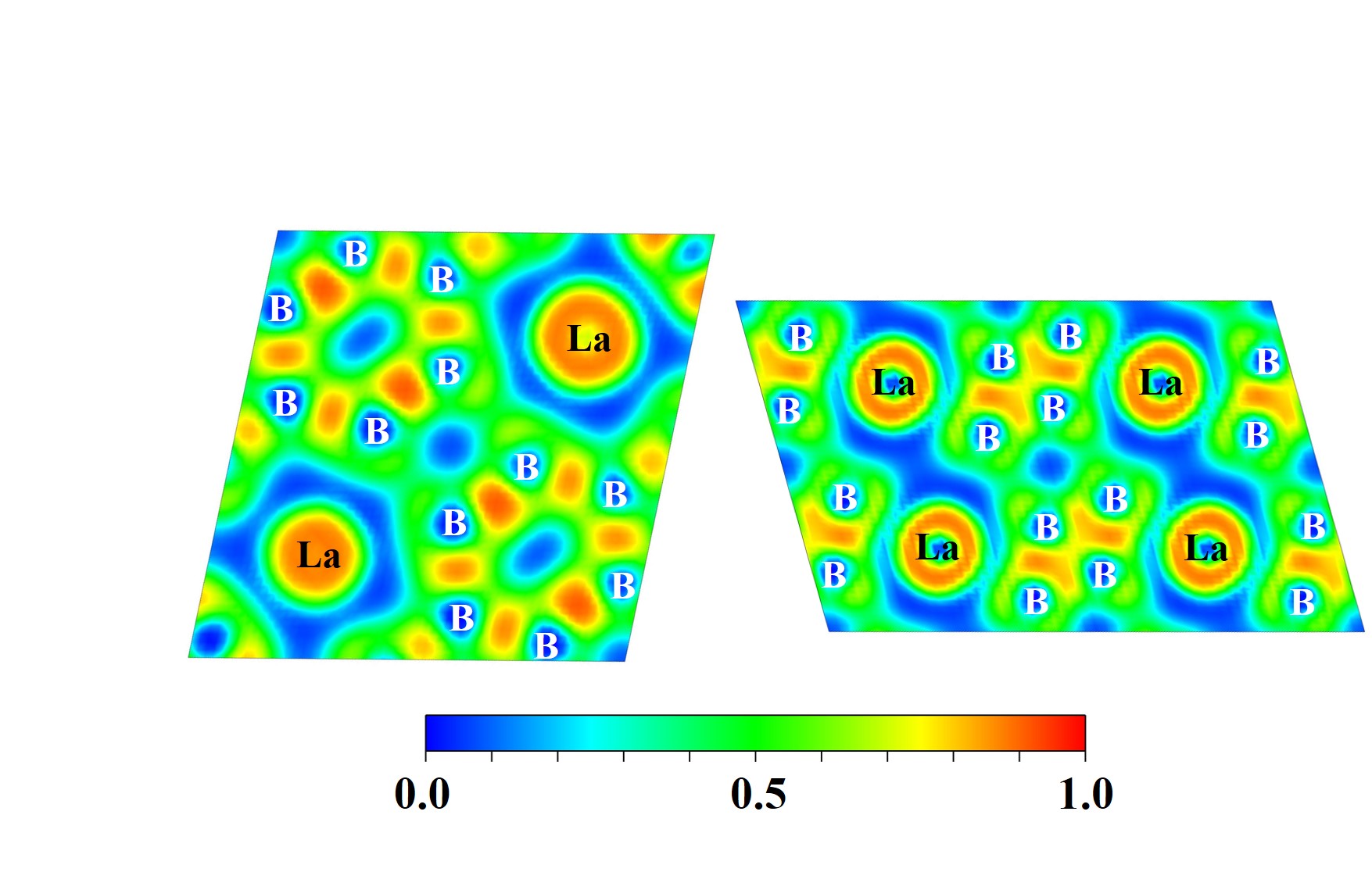}
\end{center}	
\textbf{FIG. S5.} Electron localization function (ELF) of \textit{R}-3\textit{m} LaB$_{8}$ at 100 GPa.
 \label{fig:phase}

\newpage
\begin{center}
	\epsfxsize=14cm
	\epsffile{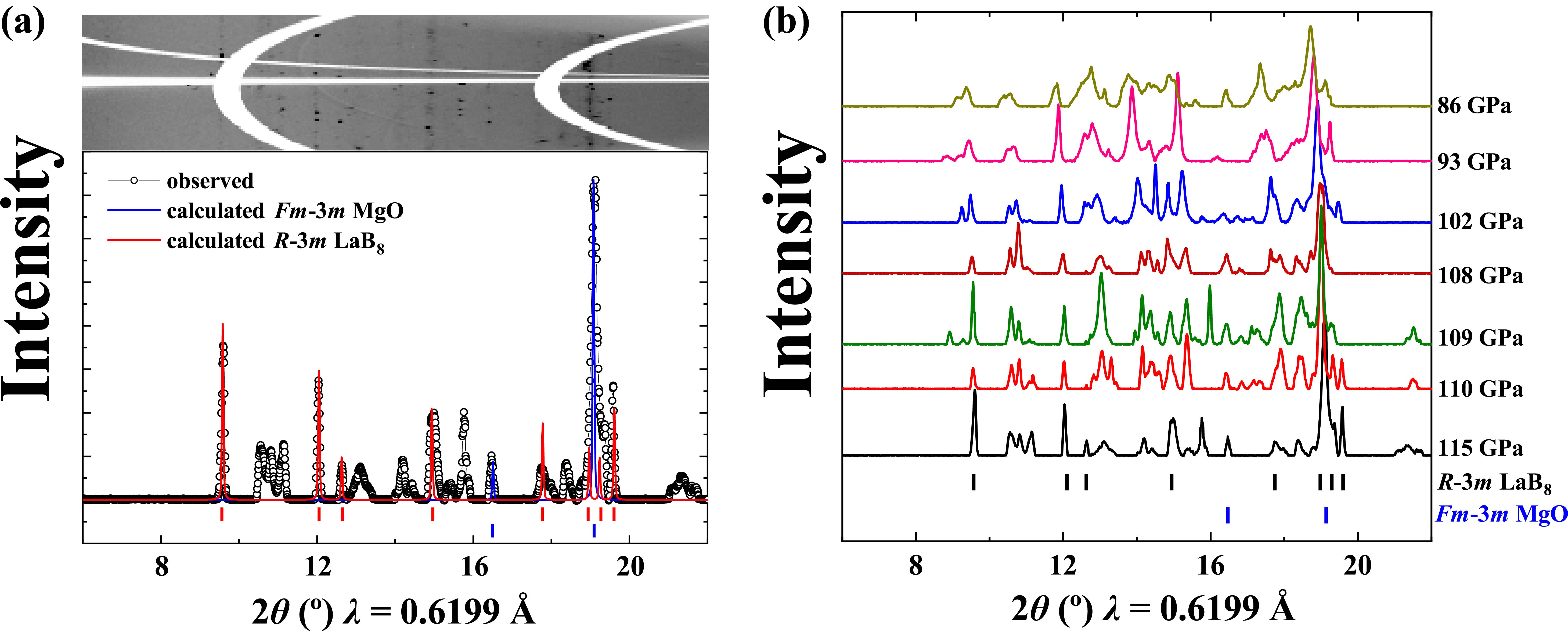}
\end{center}	
\textbf{FIG. S6.} (a) Experimental XRD patterns of cell-1 at 115 GPa. The 2D cake image is presented above the integrated pattern. Experimental data are shown as black points connected by a thin black line. Red and blue tick marks below the pattern indicate the reflections for \textit{R}-3\textit{m} LaB$_{8}$ and \textit{Fm}-3\textit{m} MgO at certain pressure. Several unidentified may be from other unconsidered stoichiometry for current studies. (b) Experimental XRD patterns of cell-1 during decompression in the pressure range of 115-86 GPa. 
\label{fig:phase}

\begin{center}
	\epsfxsize=14cm
	\epsffile{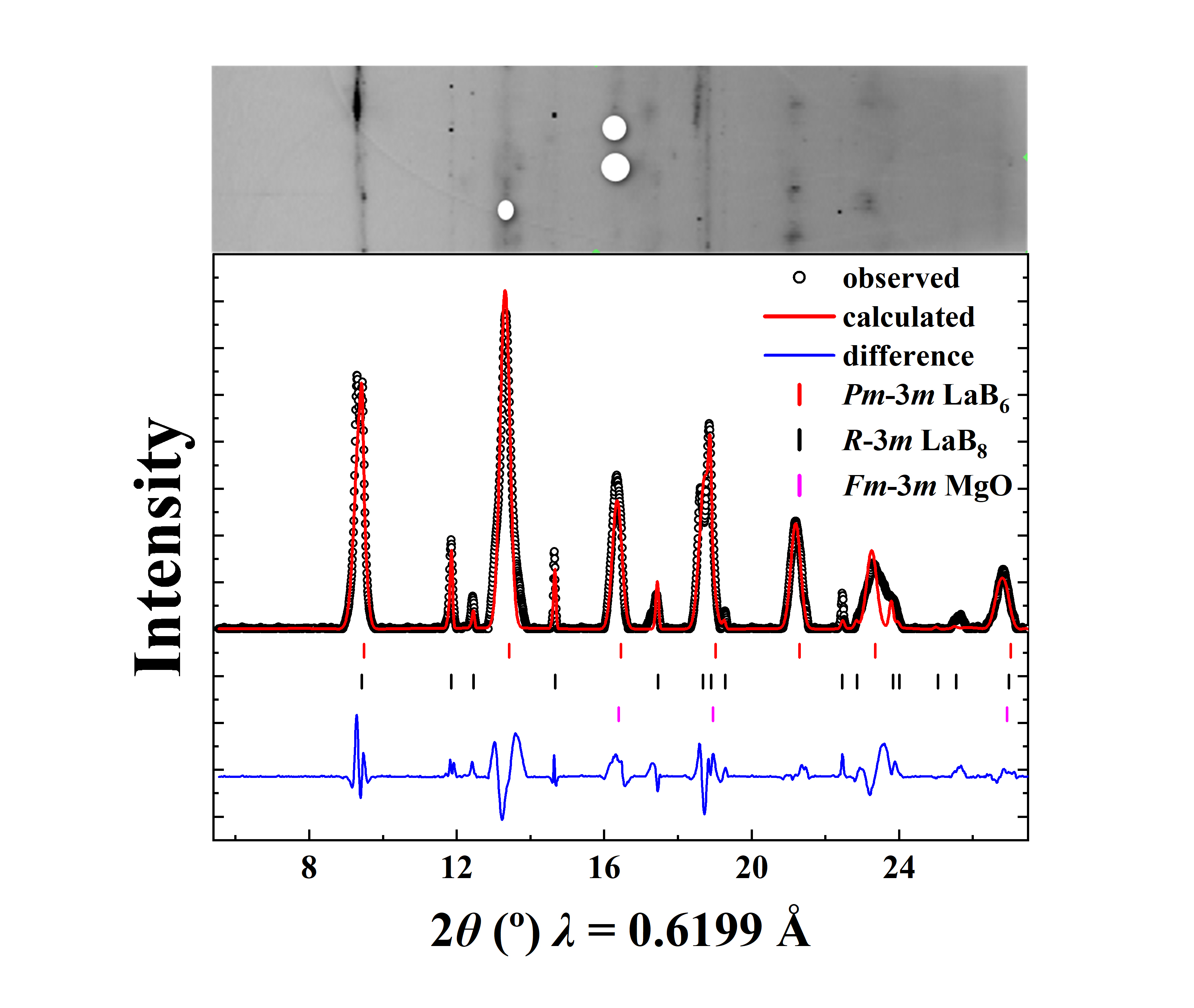}
\end{center}	
\textbf{FIG. S7.} Sample of cell-2 at 108 GPa. Experimental XRD data (black points) collected at 108 GPa with Rietveld refinement (red line). The 2D cake image is presented above the integrated pattern. Red, black and purple tick marks below the pattern indicate the reflections for \textit{Pm}-3\textit{m} LaB$_{6}$, \textit{R}-3\textit{m} LaB$_{8}$ and \textit{Fm}-3\textit{m} MgO at certain pressure. 
\label{fig:phase}

\begin{center}
	\epsfxsize=14cm
	\epsffile{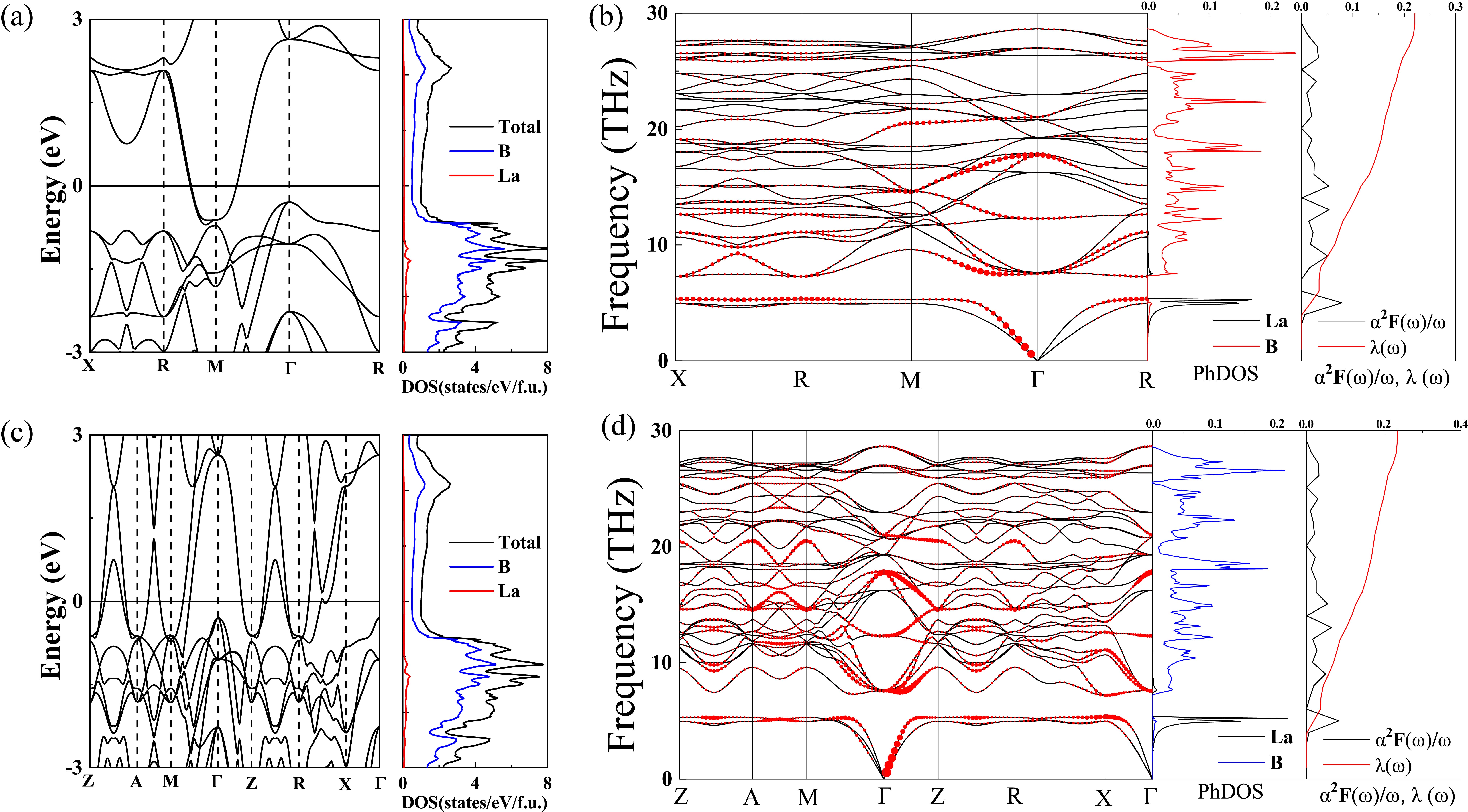}
\end{center}	
\textbf{FIG. S8.} Band structures, phonon dispersion relations, projected phonon density of states (PHDOS) and Eliashberg spectral function of (a),(b) \textit{Fm}-3\textit{m} LaB$_{12}$ and (c),(d) \textit{I}4/\textit{mmm} LaB$_{12}$.
\label{fig:phase}

\begin{center}
	\epsfxsize=14cm
	\epsffile{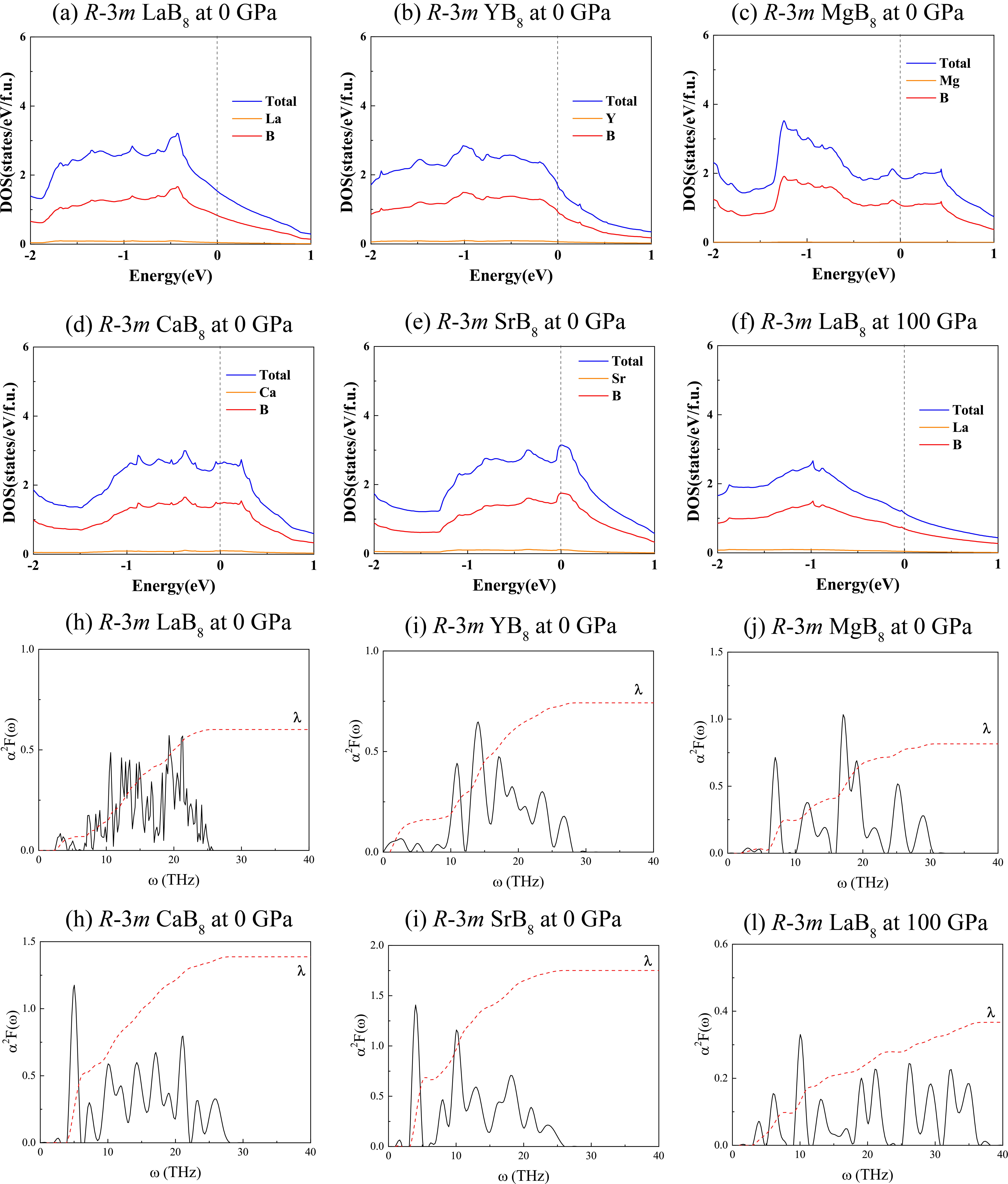}
\end{center}	
\textbf{FIG. S9.} The calculated density of states (DOS) and Eliashberg spectral function of the clathrate MB$_{8}$ compounds. 
\label{fig:phase}

\begin{center}
	\epsfxsize=14cm
	\epsffile{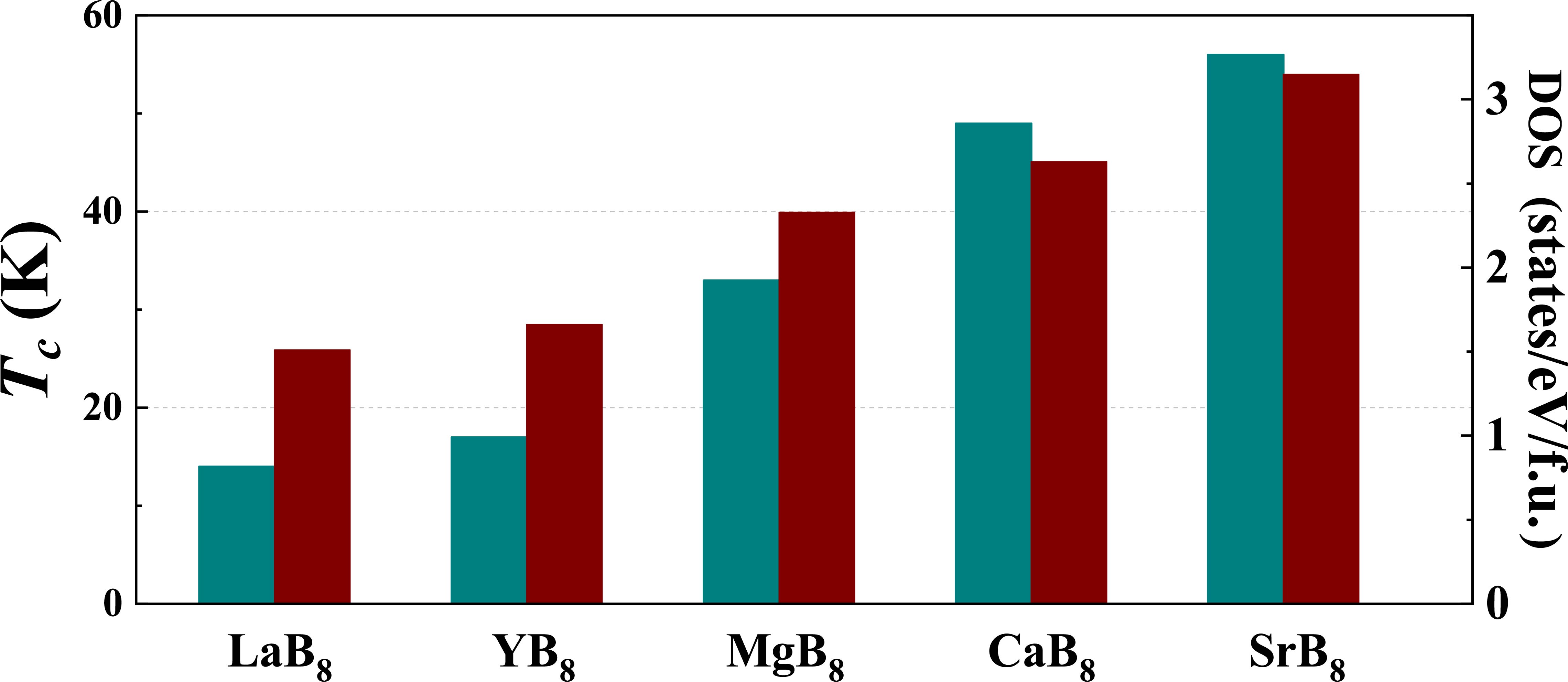}
\end{center}	
\textbf{FIG. S10.} The detailed value of \textit{T$_{c}$} (cyan bar) and density of states at the Fermi Level (wine bar) of various doped MB$_{8}$ (M = La, Y, Mg, Ca and Sr) compounds at 0 GPa. 
\label{fig:phase}

\newpage
\begin{flushleft}
	\textbf{TABLE S1.} Structural information (space group, lattice parameters and atomic coordinates) of the predicted stable La-B phases at certain pressure.
\end{flushleft}
\begin{ruledtabular}
	\begin{tabular}{lllllll}
		Phase&Pressure&&\multicolumn{4}{c} {Atomic coordinates (fractional)}\\
		(La-B sublattice)&(GPa)&Lattice Parameters (\AA,$^\circ$)&Atoms&X&Y&Z\\ 
		\hline
		\textit{R}-3\textit{m} LaB& 50 GPa & a = b = 3.069 & La (6c)& 0.000 & 0.000 & 0.102 \\
		& & c = 18.426 & B(6c)& 0.000& 0.000& 0.000 \\
		& & $\alpha$ = 90 & & & &  \\
		& & $\beta$ = 90 & & & &  \\
		& & $\gamma$ = 120 & & & & \\
		\hline
		\textit{Cmmm} LaB$_{4}$& 90 GPa & a =  15.200 & La (4j)& 0.000 & 0.253 & 0.500 \\
		& & b = 5.401 & La (4h)& 0.185& 0.500& 0.500 \\
		& & c = 3.857 & B (4g)& 0.447& 0.500& 0.000 \\
		& & $\alpha$ = 90 & B (4g)& 0.059& 0.500& 0.000 \\
		& & $\beta$ = 90 & B (8o)& -0.357& 0.500& 0.288 \\
		& & $\gamma$ = 90 & B (8p)& -0.110& 0.244& 0.000 \\
		& & & B(8p)& 0.288& 0.653& 0.000 \\
		\hline
		\textit{P}4/\textit{mmm} LaB$_{5}$& 100 GPa & a = b = 3.929 & La (1a)& 0.000 & 0.000 & 0.000 \\
		& & c = 2.775 & B (4o)& 0.795& 0.500& 0.500 \\
		& & $\alpha$ = 90 & B (1c)& 0.500& 0.500& 0.000 \\
		& & $\beta$ = 90 & & & & \\
		& & $\gamma$ = 90 & & & & \\
		\hline
		\textit{Cmmm} LaB$_{6}$ & 75 GPa & a = 8.272 & La (4a)& 0.795 & 0.000 & 0.000 \\
		& & b = 6.590 & B (4j)& 0.000& 0.881& 0.500 \\
		& & c = 3.826 & B (8n)& 0.000& 0.309& 0.206 \\
		& & $\alpha$ = 90 & B (8q)& 0.342& 0.773& 0.500 \\
		& & $\beta$ = 90 & B (4h)& 0.405& 0.000& 0.500 \\
		& & $\gamma$ = 90 & & & & \\	
		\hline
		\textit{R}-3\textit{m} LaB$_{8}$ & 75 GPa  & a = b = 4.816 & La (3a)& 0.000 & 0.000 & 0.000 \\
		& & c = 8.518 & B (6c)& 0.000& 0.000& 0.711 \\
		& & $\alpha$ = 90 & B (18h)& -0.143& -0.286& 0.567 \\
		& & $\beta$ = 90 & & & & \\
		& & $\gamma$ = 120 & & & & \\	
	\end{tabular}	
\end{ruledtabular}

\begin{flushleft}
	\textbf{TABLE S2.} Bader charge transfer of \textit{R}-3\textit{m} LaB$_{8}$ at 100 GPa.
\end{flushleft}
\begin{ruledtabular}
	\begin{tabular}{lll}
		Phase&Atoms&Charge(e)\\
		\hline
		LaB$_{8}$&La&2.60\\
		&B1 & -0.32 \\
		&B2 & -0.16 \\
	\end{tabular}	
\end{ruledtabular}

\begin{flushleft}
	\textbf{TABLE S2.} Crystal structure of experimentally synthesized \textit{R}-3\textit{m} LaB$_{8}$ phases.
\end{flushleft}
\begin{ruledtabular}
	\begin{tabular}{lllllll}
		Phase&Pressure&&\multicolumn{4}{c} {Atomic coordinates (fractional)}\\
		(La-B sublattice)&(GPa)&Lattice Parameters (\AA,$^\circ$)&Atoms&X&Y&Z\\ 
		\hline
		\textit{R}-3\textit{m} LaB$_{8}$ & 130 GPa  & a = b = 4.760 & La (3a)& 0.000 & 0.000 & 0.000 \\
		& & c = 8.404 & B (6c)& 0.000& 0.000& -0.746 \\
		& & $\alpha$ = 90 & B (18h)& -0.139& -0.279& 0.573 \\
		& & $\beta$ = 90 & & & & \\
		& & $\gamma$ = 120 & & & & \\	
	\end{tabular}	
\end{ruledtabular}